# The Compton scientific mission in Brazil in 1941: a perspective from national newspaper and documents of the time

Francisco Caruso[1], Adílio Marques[2], Felipe Silveira[3]

[1]Centro Brasileiro de Pesquisas Físicas, Rua Dr. Xavier Sigaud, 150 – 22290-180 Urca, Rio de Janeiro, RJ, Brazil

[2]Universidade Federal Rural do Rio de Janeiro, Av. Gov. Roberto Silveira, s/n - 26020-740 Moquetá, Nova Iguaçu, RJ, Brazil

[3]Independent Researcher, Amadora, Lisbon, Portugal

**ABSTRACT**

Starting from the perspective of reports published in Brazilian newspapers at the time, as well as letters exchanged between scientists who worked in Brazil and North American colleagues and documents from the symposium on cosmic rays, a chronological sequence of how the so-called *Compton mission* in Brazil took place and was perceived by the literate public will be presented.

**INTRODUCTION**

The subject of this paper is the *Compton mission* in Brazil, in 1941, and its scope is like that of a publication concerning Enrico Fermi's visit in Brazil, in 1934 [1-2], namely, to present a historical reconstruction of his visit from journalistic articles released in Brazil at that time. This approach has the appeal of leading to the perception of how the fact was seen from a political and social point of view. For this accomplishment, the archives of the main newspapers of the time reporting Compton's trip to Brazil in 1941 were investigated.

Before moving on to the central theme of this paper, it is important to highlight the prominent role that Cosmic Ray Physics had on the first steps of physical research in Brazil, since the creation of the University of São Paulo, in 1934, the Compton Mission, in 1941, pion discovery by the Bristol Group [3-4], in 1947, to the creation of the Brazilian Center for Physical Research (CBPF), in 1949, and the institutionalization of Physics in the country, which still took some time to become effective. Indeed, one can safely affirm that this continuous process begins when the Italo-Ukrainian physicist Gleb Vassielievich Wataghin [5] came to São Paulo in Fermi's place [2], where he created an important experimental research group on Cosmic Rays. The initial research was on the multiple production of mesons involved young scientists such as Marcelo Damy de Souza Santos, who was a student in the first class that Wataghin taught in Brazil. Then came Paulus Aulus Pompeia, who became his assistant, Oscar Sala, Yolande Monteux. Later, others joined the group, such as Giuseppe Occhialini [6], who came from Italy invited by Wataghin to work at USP from 1937 to 1944, and Cesar Lattes. This Cosmic Ray group had a strong scientific motivation for Compton's visit, although one cannot asseverate that it was free of political connotation and professional interests. In practice, there was a significant participation of Wataghin's group at the Cosmic Rays Symposium organized as part of the Compton Mission. Indeed, two thirds (14/21) of the Communications were presented by Brazilian scientist or foreigners who had settled in the country.

To be honest, it is not possible to separate Cosmic Ray Physics from Brazilian scientific policy and how it fits into a larger context of foreign relations during the period when Physics flourished in Brazil, between 1930-1950, and even afterwards (beginning of institutionalization) as shown in the references [7-8]. In such broad scenario, for the Compton Mission to be carried out, there was a real diplomatic battle, due to WWII [9].

**THE SCIENTIFIC MOTIVATION**

Born in 1892, the American physicist Arthur Holly Compton earned his Ph.D. from Princeton University in 1916. Following a year as a physics instructor at the University of Minnesota, he transitioned to a role as a research engineer at the Westinghouse Lamp Company in Pittsburgh. Subsequently, more precisely in 1919, he embarked on a journey to the University of Cambridge, assuming the role of a National Research Council Fellow.

In 1920, Compton moved to Washington University in St. Louis, ascending to the position of the head of the physics department. Three years later, in 1923, he furthered his academic journey at the University of Chicago, where he served as a physics professor for the next 22 years, returning to St. Louis as chancellor in 1945 and, from 1954 until his retirement in 1961, he served as a professor at the University of Washington.

The culmination reward for his groundbreaking work came in 1927 when, in collaboration with C.T.R. Wilson, Compton received the Nobel Prize in Physics. This prestigious accolade honored their discovery of the increase in the wavelength of X-rays, because of the scattering of incident X-rays by free electrons. This revelation led to the understanding that the scattered quanta possess less energy than those of the original beam, a phenomenon now universally recognized as the *Compton effect* [10, 11, 12], which was essential to put in evidence with less criticism the corpuscular nature of light.

Since the late 1920s, Compton has already joined experiments on cosmic rays in the Kashmir mountains in India and, in 1930, he had set up an experimental program with the intention to study cosmic rays at high elevations such as found in South America, confirming the observations (1927) by Jacob Clay, on voyages between Holland and Java, of the influence of latitude on cosmic ray intensity. It is important to highlight that such experiments in South America were not new at the time, since they had been carried out by the Nobel Prize winner and former Compton professor, Robert Andrews Millikan, since the beginning of the 1920s. The studies carried out by Compton ended up creating a controversy with Millikan about the nature of cosmic rays since his results differed from previous results obtained by Millikan [13].

Compton regularly participated in several international Cosmic Ray symposiums. He was aware of the controversy over the production of "heavy electrons" (initially called mesotrons and later mesons): would it be multiple, from a single collision between the cosmic rays and the atmosphere, or plural, as Werner Heisenberg had suggested in 1936-37 [14, 15]? The first alternative was based on an experiment done by Wataghin, Souza Santos e Pompeia at the Faculty of Philosophy, Sciences and Letters of the São Paulo University, with the help of Brazilian Air Force (FAB) planes went up to a heigh of 7 km, published, in 1940, in *Physical Review* [16].

In a letter dated January 4, 1941, Compton thanks Wataghin for sending him the manuscript of his article and confirms his intention to carry out new cosmic ray experiments in South America,[1] more specifically in Bolivia, Peru and following the suggestion by Paulus Aulus Pompeia, also included, in Brazil, the state of São Paulo in his experiences. In this same letter, Compton invites Wataghin to collaborate in his research [17]:

> *We know that you are considering also the carrying on of experiments on cosmic rays in the mountains. We should be very happy if it is possible to cooperate in some or all of the experiments that we have just outlined. it would be especially pleasant if you can have an expedition in the Andes that could work together with ours while we are there in June and July, or if sone one from your laboratory could collaborate with us in the balloon experiments in southern Brasil (sic.). I invited Pompeia to go with us on this expedition, but he does not find it parcticable (sic.) at this time.*
>
> *We shall keep you advised from time to time as our plans progress.*

Compton led the expedition, strategically selecting various regions in Brazil, Peru, and Bolivia due to their proximity to the magnetic equator [18].[2] In Brazil, specialized balloons equipped with measuring instruments were released, reaching altitudes of 25 kilometers. Meanwhile, in Peru, a series of experiments were conducted to capture photographic trajectories and measure the intensity of cosmic rays. This collaborative effort marked a significant contribution to the understanding of cosmic rays and their behavior in the unique geographical settings of South America during a pivotal period in history.

Professor William Polk Jesse together with the Brazilian scientist Paulus Aulus Pompeia from the University of São Paulo, who was commissioned with the expedition will be responsible for the research in Brazil. In Peru, Dr. Ernesto O. Wolan and Dr. Donald J. Hughes were responsible for photographing cosmic rays in San Cristobal near Lima. Dr. Norman Hilberry and his wife Dr. Ann Hepburn Hilberry, both from New York University, were responsible for the cosmic ray measurements [19, 20, 21].

Wataghin, together with the president of the Brazilian Academy of Sciences Arthur Moses, motivated by the trip that Compton would already take to Brazil organized an International Symposium on Cosmic Rays which took place in Rio de Janeiro, at the end of Compton's visit.[3]

---

[1] Wataghin had previously been informed in a letter sent by Pompeia, in December 1940, of Compton's intentions to pass through Brazil on his expedition [23].

[2] The intensity of the magnetic field tends to decrease from the poles to the magnetic equator. In addition, South America has a region known for the effect of the South Atlantic Anomaly (SAA) that leads to an even greater flow of energetic particles in this region [24, 25].

In a further letter written by Compton to Wataghin, on April 21, 1941, already aware of the symposium, Compton requests that it take place between the 4th and 6th of August [22]:

> *Mr. Pompeia has told us of your plans with regards (sic.) to a symposium at Rio de Janeiro. It will be most convenient for us if this could be neld (sic.) from August 4 to 6. This would let us complete the work in Peru in time for the men there to come to Rio for the conference. Immediately afterwards most of us would return to the United States.*

However, Compton's arrival in Brazil was not only motivated by findings published by Brazilian scientists; it also occurs amidst a geopolitical backdrop where the United States sought alignment with different South American countries in support of the Allied forces during World War II [9]. The International Symposium took place at Rio de Janeiro between August 4$^{th}$ and 8$^{th}$ 1941, and its proceedings were published two years later [27].

**THE COMPTON MISSION**

On June 18, 1941, Pompeia was the first member of the Compton mission arrived in Rio de Janeiro aboard the transatlantic Argentina. Pompeia attended, during the period from 1940 to 1942, to improvement course at the Compton laboratory in Chicago.[4] And together with Gleb Wataghin, carried out preliminary experiments on the arrival of the Compton mission in Brazil [28].

On the July 16th, from the municipality of Jaú, located in the central region of the state of São Paulo, the meteorology service of the Ministry of Agriculture in Rio de Janeiro, under the leadership of meteorologist José Carlos Junqueira Schmidt, conducted the first set of a series of aerological soundings. The objective was to facilitate the future retrieval of instruments designated for experiments carried out by the Compton mission. Aimed to determine the optimal location for the official launch of the instrument. In this initial sounding, the balloons landed near the city of Bragança, located 221 km southeast of Jaú [29, 30, 31]. On the same day, Jesse, aboard the ship Uruguay, passed through Rio de Janeiro and departed the following day for Santos, [32, 33, 34]. On board the ship, Jesse was interviewed by the newspaper *O Globo* [35], when he clearly defined the goal and the strategy of the mission, declaring:

> *Cosmic rays come from regions beyond our planetary system. Upon reaching the great heights of the Earth's atmosphere, they form very interesting particles called "mesotrons", which are characterized by*

---

[3] In addition to Compton, Wataghin expressed in a letter to Moses the intention of inviting physicists like Enrico Fermi, Hans Bethe, Bruno Rossi and Wolfgang Pauli to the symposium [26].

[4] Some report at the time mistakenly identified him as a North American scientist and in other reports his name appears as Pandos Pompéia.

> *their great penetrating power. In the depths of the mines, "mesotrons"[5] have been measured that cross 400 meters of solid rock.*
>
> *The objective of our expedition is to determine the number of "mesotrons" formed in the high atmospheric regions here in Brazil, close to the earth's magnetic equator, compared with the number already found in Chicago.*
>
> *For this purpose. A measuring device of "mesotrons" is dragged by 15 to 20 balloons loaded with hydrogen. At a height of 25 kilometers some of these balloons burst and the device slowly falls to earth supported by the other balloons that remain intact. When the Device is retrieved, a photographic film reveals the altitude reached, the temperature and the intensity of the "mesotrons" during the entire flight, which lasts approximately eight hours and covers approximately 150 kilometers.*

On July 18th and 21st, Schmidt, continuing his sounding efforts, launched two additional sets of balloons that reached the city of Pouso Alegre, in the state of Minas Gerais. Pouso Alegre is located 271 km east of Jaú, and the second wave of balloons landed in the vicinity of the municipality of Itapira, approximately 180 km east of Jaú. As a result of these soundings, the city of Bauru, located 50 km west of Jaú, was chosen as the most favorable location to proceed with the experiments [30, 36].

On the July 27th, Compton arrived in São Paulo where he was received by few members of the Brazilian Academy of Sciences: Alvaro Alberto da Mota e Silva, Joaquim da Costa Ribeiro, Guilherme Florence, and Francisco Emygdio da Fonseca Telles and then was taken to the Hotel Esplanada [37, 38, 39].[6]

Between the 28th and 31st of July, devices designed to record cosmic radiation, shown in fig. 1 and fig. 2, were launched at altitudes of up to 20 or 30 kilometers into the stratosphere. Using the data obtained in the test launches from the 16th to the 21st of July, four cities in the state of São Paulo were chosen for the official launch: São Pedro de Piracicaba, Jaú, Baurú, and Marília. In total, 11 groups of devices were launched, dragged by hydrogen balloons, fig. 3, 4 and 5 [27].

---

[5] Discovered by the American scientist Carl D. Anderson in 1936, the name *mesotron* was coined by him and Seth Neddermeyer in a letter to *Nature*, in 1938, composed by the prefix "meso", which in Greek means *mid*, as its mass is between the mass of the electron and that of the proton. It was later renamed *muon*.
[6] Building where the headquarters of the Secretary of Agriculture and Supply of the State of São Paulo are located today.

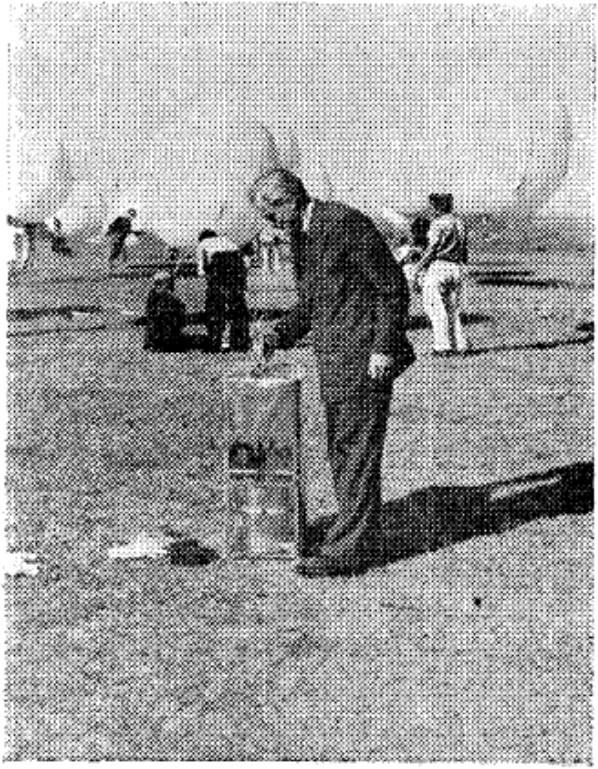

*Figure 1: Arthur H. Compton next to the devices a moment before the launch of the balloons was about to begin.*

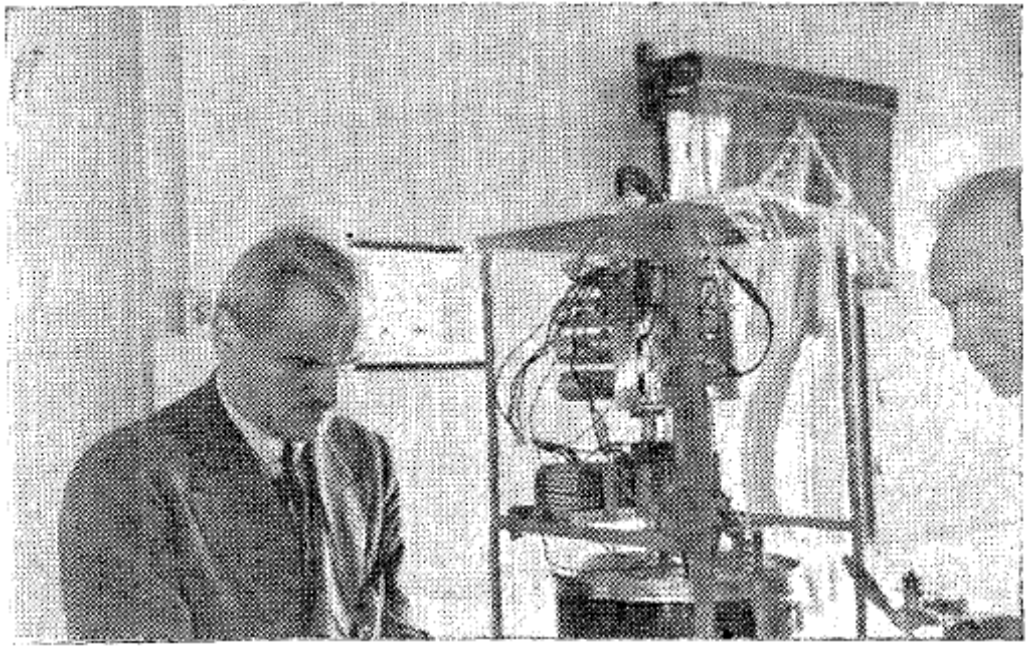

*Figure 2: Arthur H. Compton next to the devices before the launch of the sounding balloons was about to begin.*

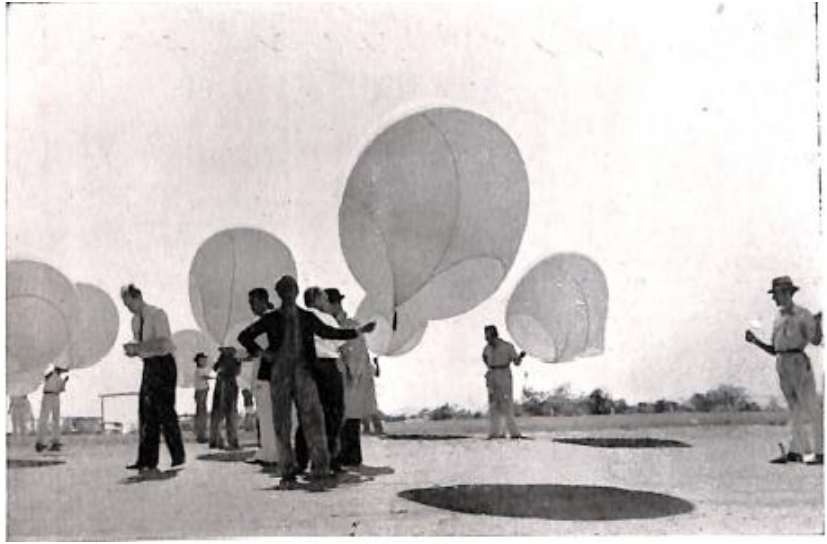

*Figure 3: Arthur H. Compton and William Jesse preparing the devices at Baurú's airport.*

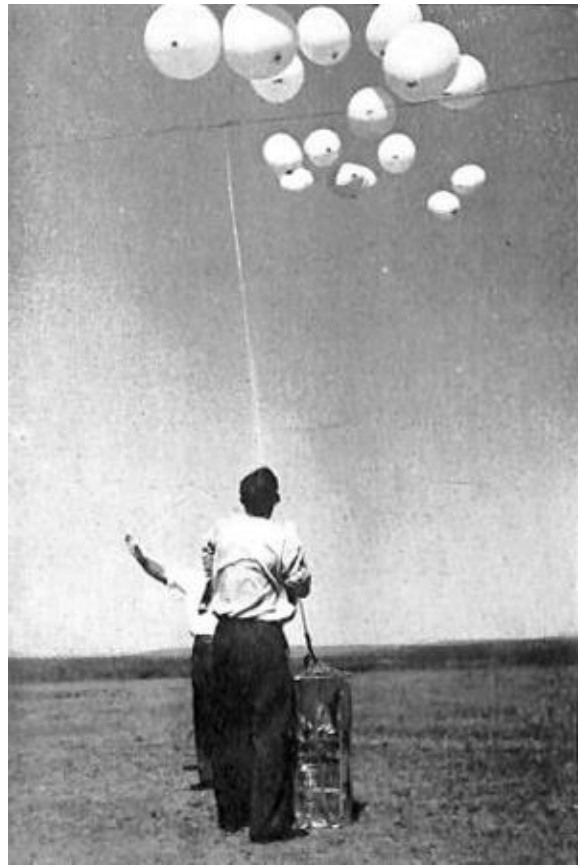

*Figure 4: Devices being launched dragged by hydrogen balloons.*

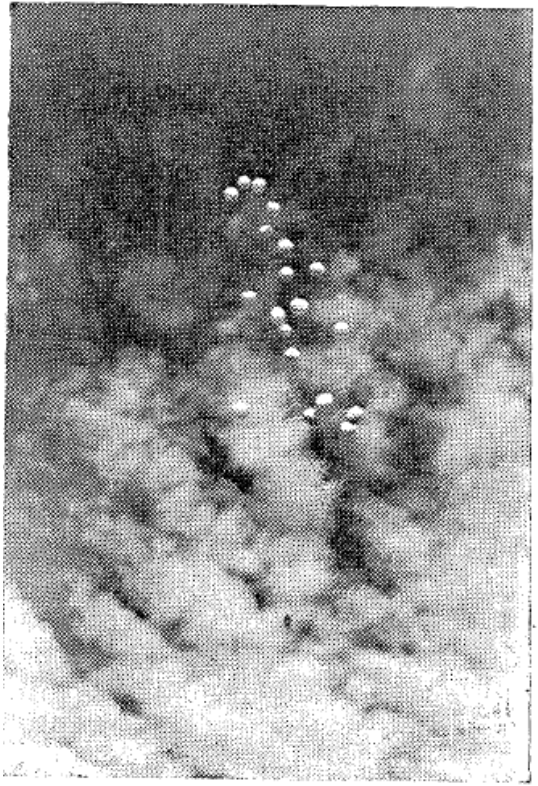

*Figure 5: Set of 16 balloons carrying the recording devices.*

In the edition of July 31, 1941, Compton gives an interview to the Jornal *A Noite* [40] detailing what the experiments are about and anticipating the expected conclusions:

> *The experiments consist of observations made with recording devices, which are dragged by balloons to high altitudes. Weighing approximately 10 kilos, these devices can reach heights of over 25 kilometers, where the rays coming from outside are not greatly affected by the layer of air crossed. The recent measures taken in the USA. By means of balloons and similar devices. They demonstrate that the particles that penetrate the atmosphere are composed of protons. These constitute the heavy part of hydrogen atoms containing a positive electrical charge.*
>
> *If this assertion is rigorously proved in the vicinity of the magnetic equator, we will conclude that only high energy protons can reach the atmosphere and that they come more frequently from the West. Particularly interesting will be the study of energy transmission from protons to the secondary rays they produce.*
>
> *For this study it will be necessary to keep the balloons in the air for many hours, therefore subject to the winds. Therefore, they can fall many tens of kilometers from the starting point. Only with the recovery of the balloons and the respective equipment, the results of the experiments can be obtained. In some cases, these balloons may be accompanied by an airplane. With the cooperation of the Brazilian government, the physics department of the same university, and the Brazilian press, the expedition hopes to be able to recover the balloons that carry a label declaring the prize and the address to which whoever finds them should go.*

> *Messrs. Hugues and Wollan will photograph the trajectories of cosmic rays. These photographs will be obtained by means of condensing ions that are visible and photographable and that will reveal the type of particle, the energy value, and the direction in which it moves. At the accessible altitudes of Peru, types of particles rarely observed at sea level are likely to be found.*
>
> *It is also hoped that with the studies of Mr. Hilberry in "El Misti", it is possible to know how many particles there are of a certain energy produced by the cosmic rays that reach the atmosphere. Its devices, including thermionic valves and batteries, will need to be transported on donkeys through trails, to climb the peak of the mountain, at an altitude of approximately six thousand meters.*
>
> *The study of cosmic radiation has become one of the most active fields of research: it is the best available method for investigating the fundamental particles that make up matter. Crystals are composed of molecules, molecules are composed of atoms, which in turn are composed of parts known as electrons, protons, and neutrons, etc.*
>
> *With cosmic rays, the relationships between these various elementary particles can be studied. The previously unknown positive electron has been found in cosmic rays. The same happened with the mesotrons, which have a mass of intermediate value between the mass of the electron and that of the proton. By means of cosmic rays it is possible to study the transformation of protons into mesotrons and of these into electrons. The basic investigation of matter is scientifically as important as its study at other levels, such as the study of the distribution of atoms that form molecules and which we call chemistry. So far, its practical importance has not become evident. However, it seeks to increase the basic knowledge of the fundamental structure of matter.*

On August 1st, at 9 pm, Professor Compton gave a lecture on the theme Cosmic Rays, in the main hall of the Faculty of Philosophy, Sciences and Letters, in São Paulo. For more than two hours, Compton made a complete and detailed presentation of his research in the Americas. He used photographic projections for illustrating his conference [41, 42, 43].

On the 3rd of August, Compton, together with his assistants and their respective wives, arrived in Rio de Janeiro from São Paulo by the Cruzeiro do Sul train [44, 45, 46, 47].[7] He was received at the Pedro II station by members of the Brazilian Academy of Sciences and accompanied by them to the Hotel Gloria,[8] where they will be staying. On that day, the *Grande Prêmio Brasil* was held at the Gavea Hippodrome, where the Compton mission together with the Portuguese writer António Joaquim Tavares Ferro, known as António Ferro, was received for a lunch hosted by the journalist and politician Lourival Fontes, Director of the press and advertising department (DIP).[9]

---

[7] Also known as Expresso Cruzeiro do Sul, it connected the cities of Rio de Janeiro and São Paulo, during the years 1929 to 1950, by the central railroad of Brazil.

[8] Luxury hotel located in the neighborhood of Glória, in Rio de Janeiro. Room 400 hosted Albert Einstein, for a week in May 1925, during his visit to the city. The Hotel remained open for 86 years until it closed in 2008. The building is currently undergoing renovations to become a residential building.

[9] Created in 1939 by then President Getúlio Vargas, it served as a propaganda and government censorship instrument during the Estado Novo. In 1945 it was replaced by the National Information Bureau.

On the 4th of August, Compton was interviewed by the newspaper *A Noite* [48] telling a little about his experiences in São Paulo:

> *The experiments carried out show not only the capacity of Brazilian scientists, but also allowed me to assess the kindness and hospitality of their country. The experiments in São Paulo were very successful.*

Continuing with the interview, Compton commented a little about what he would say in his presentation.

> *In my communication today, I will address Alf'ven's theory on the origin of cosmic rays and also, I will address the magnetic fields of galaxies. This theory, consider the emission of electric particles in a great mass of animated stars in spiral motion.*

And at 9 am in the main hall of the National School of Engineering,[10] Compton started the opening conference, fig. 6, of a week of debates on cosmic rays organized by the Brazilian Academy of Sciences [49, 50, 51, 52, 53, 54].[11]

---

[10] Today is the Institute of Philosophy and Social Sciences of the Federal University of Rio de Janeiro.
[11] Jornal do Comércio, a Brazilian periodical, published the minutes of the Meeting of the Brazilian Academy of Sciences on the Cosmic Ray Seminar in its edition on the 16th of August, 1941, according to the collection of the University of São Paulo. In: http://acervo.if.usp.br/index.php/recorte-do-jornal-do-comercio.

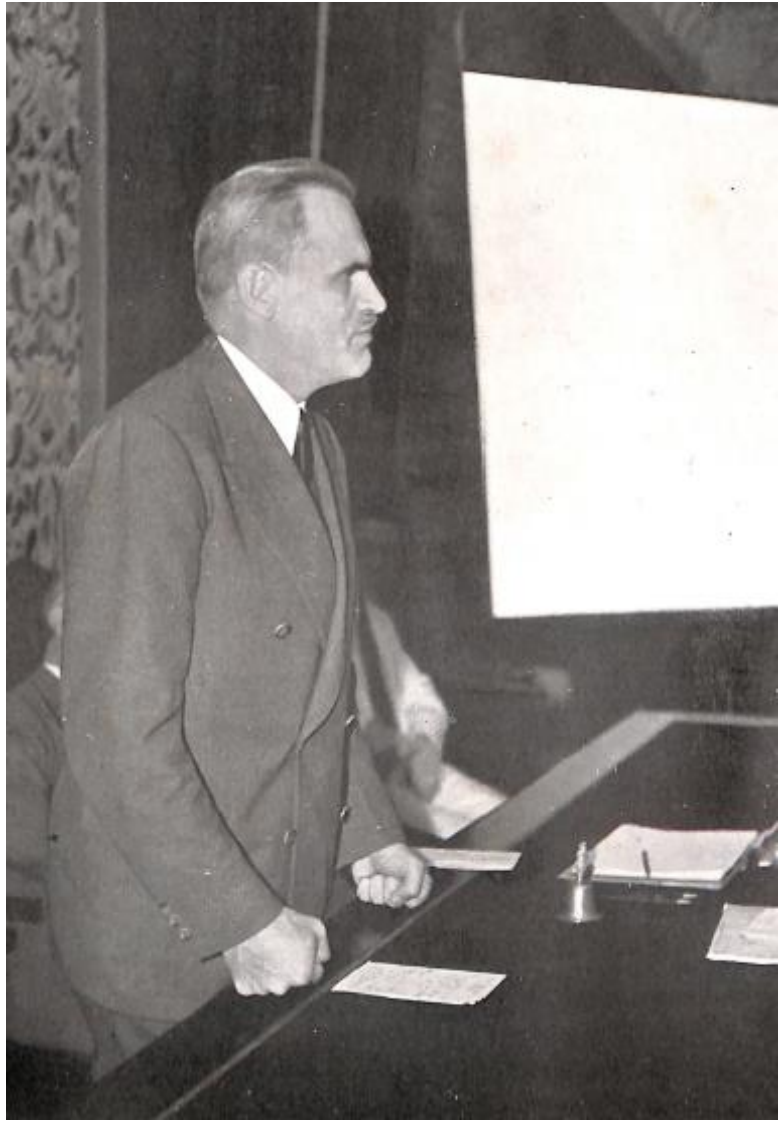

*Figure 6: Arthur H. Compton in the "Symposium sôbre Raios Cósmicos" on August 4, 1941.*

In addition to Alfvén theory,[12] he also addressed issues such as measurements of magnetic fields and cosmic radiation, including the deviation of the trajectory of cosmic rays by the magnetic mass of the Earth and other celestial bodies. Then, he passed the word to one of his assistants, William Jesse, who began his communication.

On the same day, August 4, the Minister of Foreign Affairs Osvaldo Aranha hosted a lunch at the Jockey Club for members of the Compton scientific mission. Also in attendance were the US Ambassador Jefferson Caffery, the General Secretary of the Ministry of Foreign Affairs Mauricio Nabuco, the President of the Brazilian Commission on Intellectual Cooperation Miguel Osorio de Almeida, the Rector of the University of Brazil[13] Raul Leitão da Cunha, in addition to professors Arthur Moses, Carlos Chagas, Meneses Oliva, Flexa Ribeiro and Inacio Amaral. In the afternoon, Professor Arthur Moses and his wife offered the members of the mission a reception at their residence located on Rua Muniz Barreto in Botafogo neighborhood, which was also attended by figures from the diplomatic corps of the government and society in the city of Rio de Janeiro [55, 56, 57, 58, 59, 60, 61].

---

[12] Named after its proposer, the Nobel Laurent Hannes Olof Gösta Alfvén.
[13] Current Federal University of Rio de Janeiro.

On the 5th of August, the newspaper *O Globo* reported Compton's decision to postpone his return to the United States, which would have taken place on the 7th, to the 9th of August, at the request of the *Instituto de Estudos Brasileiros* so that he could participate in another conference on the 8th [62].

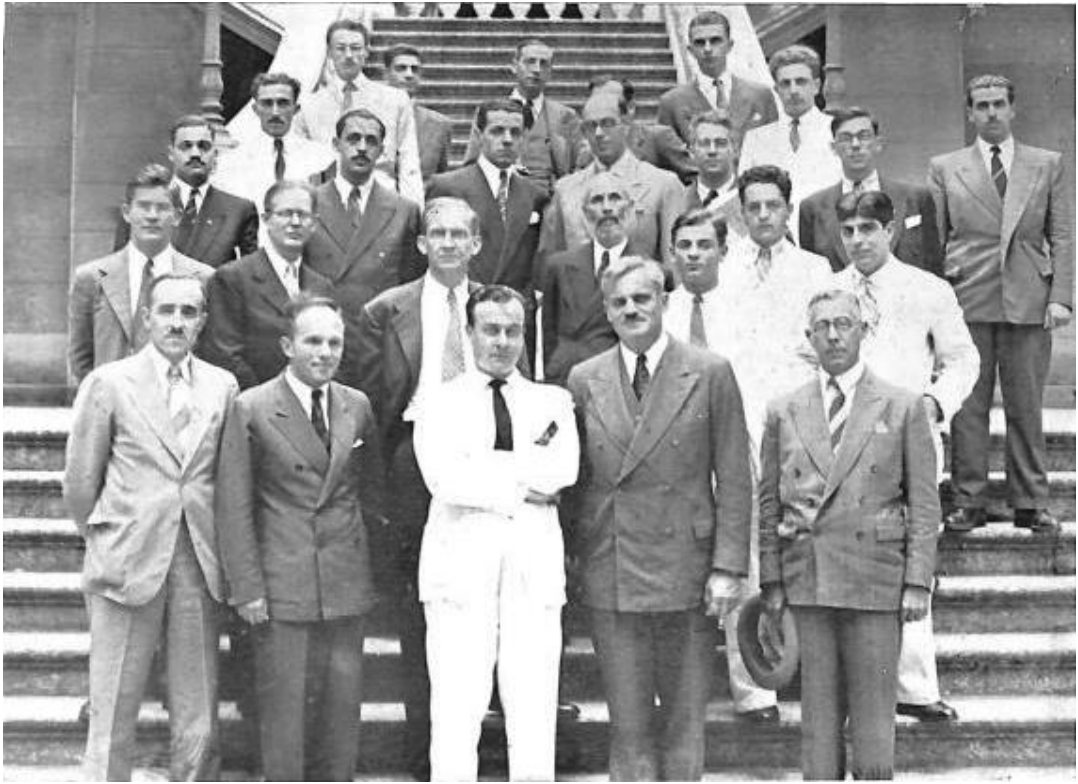

*Figure 7: participants of the "Symposium sôbre Raios Cósmicos" on August 6, 1941*

On August 6th, Compton made a night visit to the studios of radio division of DIP (*Departamento de Imprensa e Propaganda*, in Portuguese), and spoke into the microphone on a national radio program called *Hora do Brasil* [63, 64, 65, 66, 67].[14] At this opportunity, he thanks the cooperation of Brazilian physicists, essential for the achievements of the Mission, and conclude urging rapprochement between Brazil and the USA in the face of the War in Europe. In his words:

> *Listener friends. I am pleased to greet our Brazilian friends, who treated us so cordially during our brief visit. In response to the kind invitation of the Brazilian Academy of Sciences and the physicists of São Paulo and Rio de Janeiro, a small group of scientists from the University of Chicago came to your country to study cosmic rays and discuss some of our mutual problems with Brazilian scientists.*
>
> *Our experiments here have consisted of raising balloons with instruments that record cosmic rays as they approach the earth. Your*

---

[14] radio program broadcast obligatorily by all broadcasters in the country, between 7:00 pm and 8:00 pm, only with the disclosure of the acts of the Executive Branch of the federal government. It is currently called the *A Voz do Brasil* and has greater schedule flexibility.

*meteorological service and your air service have provided us with the greatest cooperation and assistance to carry out our balloon flights. Your radio and your press have also helped us a lot, keeping the public alert in search of the instruments as they descend. Even more important has been the cordial cooperation of physicists from your universities, headed by Professor Wataghin from São Paulo. Without this help, our work in Brazil would not have been possible. But, with your cooperation, we have already managed to achieve an important part of our objectives and successfully address what is missing. Above all, we took enormous advantage of the conferences we had last week with scientists from Brazil. The development of ideas arising from discussions of this nature is the most important result of a visit such as ours. We would also like to say how much we have appreciated your most cordial hospitality and the countless courtesies with which we have been showered, officially and by personal friends. Your country, like ours, faces new responsibilities. For many centuries now, we have been turning our eyes eastwards, to the countries that preceded us, seeking there our culture and our science. Gradually the strength of our civilization grew. Brazil became the leader of Portuguese culture. The United States has developed an improved industry. Now Europe is involved in a tragic struggle.*

*If the torch of civilization is to continue to shine, it is our responsibility to keep it burning.*

*We pray, with sincere hope, that Europe will quickly return to the enjoyment of its inheritance. We want to pay you our big debt. However, it is becoming more and more necessary for your country and our country to look to each other for friendship and cultural stimulation. We would like to thank you for the frank hospitality with which you received us, who seek to work with you in a common task.*

On the 8th of August, at the National School of Engineering, the last section of the symposium was held under the presidency of Compton [68, 69, 70]. At 6:30 am on August 9th, 1941, Professor Arthur Compton, accompanied by his wife, returned to the United States aboard the Douglas plane operated by the Panair company [71, 72, 73, 74, 75, 76, 77, 78].[15]

On August 21st, 1941, continuing the studies initiated by the Compton mission, at 9 am, a group of 20 balloons with cosmic ray recording devices were released in the municipality of Marilia in São Paulo. The balloons must land within the state of São Paulo and the south of Minas. People who find the balloons will be awarded prizes of 200$000[16] in cash and a medal, once they communicate the time and place to the nearest police station or the São Paulo radio patrol [79, 80, 81, 82].

On August 25th, William P. Jesse, together with Gleb Wataghin, met with federal intervenor Fernando Costa with the aim of thanking and saying goodbye, as he would then return to the United States [83].

---

[15] Leading Brazilian airline between the 1930s and 1950s.
[16] The currency of the time was the Real (popularly known as réis), a currency which still portrayed the times of colonial Brazil, was replaced by the Cruzeiro in 1942. It is estimated that 1 real was worth the equivalent of the current exchange rate in 2023 of 0.0061 US dollars.

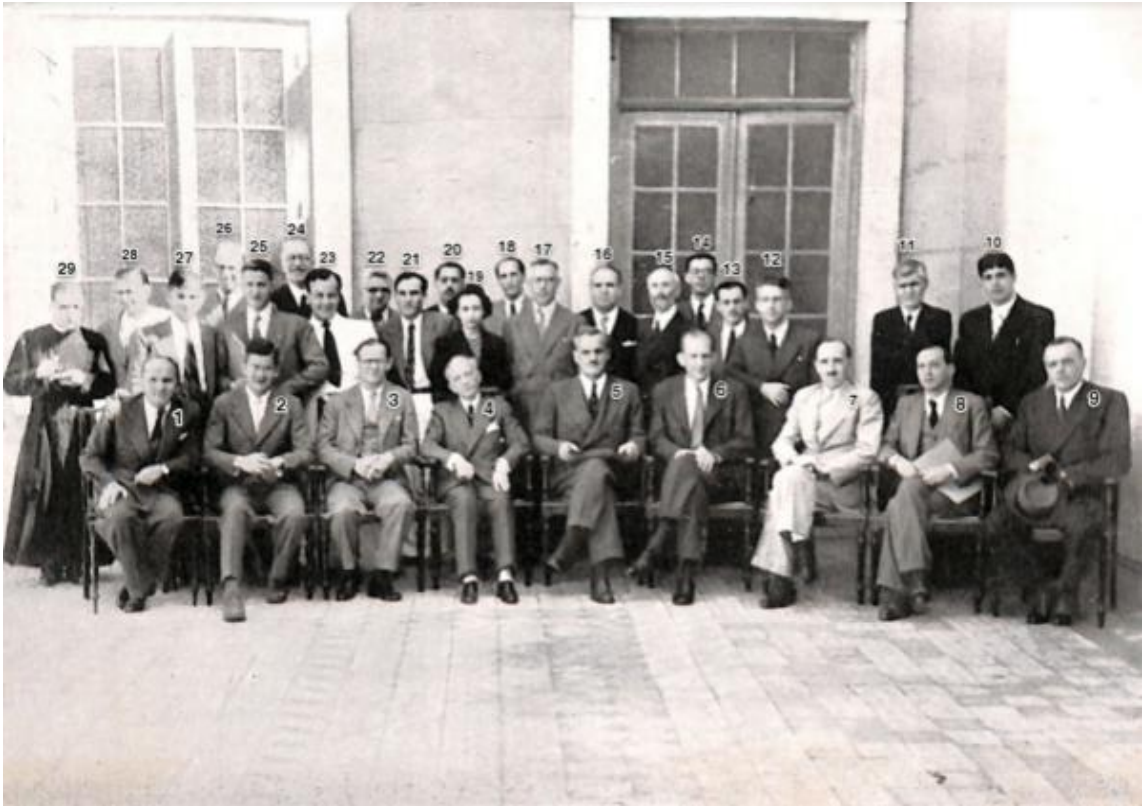

*Figure 8: Final reunion of the "Symposium sôbre Raios Cósmicos" on August 8, 1941 (1 - Gleb Wataghin; 2 - Donald Hughes; 3 - Norman Hilberry; 4 - Arthur Moses; 5 - Arthur H. Compton; 6 - William P. Lesse; 7 - Ernest O. Wollan; 8 - René Wurmser; 9 - Francisco Souza; 10 - F. M. de Oliveira Castro; 11 - F. Venancio Filho; 12 - J. Costa Ribeiro; 13 - Othon Nogueira; 14 - F. Magalhães Gomes; 15 - Arthur do Prado; 16 - Alvaro Alberto; 17 - Menezes de Oliveira; 18 - Junqueira Schmidt; 19 - Yolande Monteux; 20 - Paulo R. Arruda; 21 - Giuseppe Occhialini; 22 - M. Cruz; 23 - Carlos Chagas Jr.; 24 - Ignacio Azevedo Amaral; 25 - M. D. de Souza Santos; 26 - Bernard Gross; 27 - Abrahão de Morais; 28 - Paulus Aulus Pompeia; 29 - Pe. F. X. Roser S. .)*

On August 27, Compton, already in Chicago, made the following statements about the countries he visited on his mission [84]:

> *They correspond broadly to all gestures and cultural initiatives in the United States and, moreover, offer magnificent opportunities and complete reciprocity for any scientist.*
>
> *Equally important was the clear evidence of a great interest and an intense desire for perfect scientific and cultural cooperation.*
>
> *One of the highlights of our activities was our last week in South America, while we stayed in Rio de Janeiro. There we were, under the auspices of the Brazilian Academy of Sciences, where we met a chosen group of "physicists" and scholars, before whom we gave some lectures that were accompanied and commented by about fifty or more personalities with many of whom we were able to exchange ideas, for four days, on "solid" scientific subjects and discussing reciprocal observations on cosmic rays and related problems.*
>
> *Both for us all and for our Brazilian colleagues, these conferences and discussions were the source of inspiration for more and better scientific research.*
>
> *It was also a great stimulus for us to find in Peru such intense expressions of cultural interest, both on the part of intellectuals and academics as well as businessmen themselves.*

On October 2, 1941, a letter written by William P. Jesse [85] was published in the São Paulo Post Office.

> *Your Excellency Dr. Fernando Costa, DD. Federal intervener in the state of São Paulo, − Excellency: As we leave Brazil, I wish to express to Your Excellency Personally and to the government of São Paulo, my best thanks for the great cooperation that was given to us in the performance of our scientific mission. We received all the necessary facilities to carry out our work. I ask Your Excellency Receive, for this kindness, our warmest thanks.*

**FINAL COMMENTS**

The press material to which we had access (127 different news published in 14 newspapers), in addition to the letters between Compton and Wataghin, for example, shows that a significant space was given to Compton scientific mission, in 1941, by Brazilian written media, equivalent to what happened in 1934, with Fermi's trip to Brazil. Only Einstein's visit in 1925 deserved more attention [86].

It is true that the United States support to the expedition of a North scientific leader as Arthur Compton to Latin America was part of their policy of cultural proximity and their efforts to neutralize the influence of the Axis powers in this part of the Americas [9]. There was a similar scope in Fermi's visit, with which the Italian fascist government sought to gain sympathy in South America, in a period when Brazilian science was still very incipient. But it is also true that it contributes to the consolidation of Cosmic Ray Physics in Brazil. We can't forget that Cesar Lattes was a brilliant student in Wataghin's group and, in 1947, participate in the pion discovery [87]. From the point of view of Brazilian scientific polices, a crucial milestone.